\documentclass[11pt]{article}

\usepackage{graphicx}

\usepackage{graphicx,psfrag}

\def \s {\sigma}

\def \l {\lambda}

\def \vep {\varepsilon}

\def \a {\alpha}

\def \T {{\hbox {Tr}}}

\def \I {{\hbox{i}}}

\def \b {\beta}

\def \E {{\bf E\, }}

\def \i {{\hbox{ i}}}

\def \D {\Delta}

\def \d {\delta}

\author{A.Khorunzhy\\
D\'epartement de Math\'ematiques \\
Universit\'e de Versailles -- Saint-Quentin\\ Versailles, FRANCE
}

\begin{document}

\title{ON SPECTRAL NORM OF LARGE BAND RANDOM MATRICES}

\date{}

\maketitle

\begin{abstract}
We consider the ensemble of $N\times N$  
hermitian random matrices $H^{(N,b)}$
whose entries are equal to zero outside of the band of width $b$
along the principal diagonal.  Inside this band the entries 
$\{ H_{ij},
\ i\le j\}$  are given by independent identically
distributed gaussian random variables with zero mean value
and variance $v^2/b$.

We study asymptotic behavior of the spectral norm   
$\Vert H^{(N,b)}\Vert $ in  the limit $N\to\infty$  when the band width $b$ is
much  smaller than the matrix size $N$ 
  but also tends to infinity. 
Our main result is that if $b/(\log N)^{3}\to\infty$, then $\limsup_{N,b}\Vert 
H^{(N,b)}\Vert$ is bounded by $2v$ with probability 1. To prove this, we
derive  a  system of  recurrent relations  for the moments $M_{2k}^{(N,b)}$
and analize these relations in the limit when $k$ goes to infinity as
$N\to\infty$.
\end{abstract}

\section{ Random matrices and eigenvalue distribution}

Random matrices of infinitely increasing dimensions $N\to\infty$ were
considered first  in theoretical physics.
Physics  serves as the source of
new questions and results here and the development of
the random matrix theory reveals  many relations  between 
 random matrices and other branches of mathematics
(see the special issue on physics and mathematics of random matrix theory of
the {\it Journal of Physics A: Mathematical and General}, vol. 36 (2003)). 
In these studies, the special attention is paid to the extreme eigenvalues of
$N\times N$  hermitian random matrices with unitary invariant probability
distribution. Recent progress in this field relates
random matrix theory with  nonlinear differential equations and integrable
systems  (the pioneering mathematical results are obtained in \cite{TW}); 
from another hand 
the
probability law of the maximal eigenvalue of random hermitian matrices
$\l_{\max}^{(N)}$ is asymptotically equivalent to that of the longest
increasing  subsequence in the random permutation of
$1,\dots,N$ (see e.g.
\cite{BDK,J}). These results motivate extended studies of extreme values of
random matrix ensembles.

In present paper we study asymptotic behavior of 
$\l_{\max}$
of the ensemble of band random matrices. This ensemble 
plays an important role in the quantum chaos theory (see e.g. \cite{H}).
The probability distribution of this ensemble is not invariant
with respect to the unitary transformations.

\subsection{Wigner ensemble and the semicircle law}

The spectral theory of random matrices was started half a century ago 
by E. Wigner (see e.g. \cite{W}). 
He studied the eigenvalue distribution of the ensemble of $N\times N$ 
real symmetric random matrices $W^{(N)}$ of the form 
$$
\left( W^{(N)}\right)_{ij} = {1\over \sqrt N} w_{ij},
\eqno (1.1)
$$
where $\{w_{ij}, i\le j\}$ are independent random variables.
The main proposition proved in \cite{W} is that if $w_{ij}$ are
centered random variables of the variance
$v^2$, odd moments zero and all even moments finite,
then the normalized eigenvalue counting
function of $W^{(N)}$ defined by the formula
$$
\s_N(\l) = \#\{ \l_j^{(N)} \le \l\} N^{-1},
$$
where $\l_1^{(N)}\le \dots \le \l_N^{(N)}$ are eigenvalues of $W^{(N)}$, weakly
converges in average to the limiting function
$
\s(\l;2v) 
$ 
$$
\lim_{N\to\infty} \s_N(\l) = \s(\l;2v)
\eqno (1.2)
$$
with the density of the semicircle form;
$$
\s'(\l;2v) = \rho(\l;2v) = 
{1\over 2\pi v^2} \cases{ \sqrt {4v^2 -\l^2}, 
& if $\vert \l\vert \le 2v$,\cr
0, & otherwise. \cr }
\eqno (1.3)
$$
This statement is known as the semicircle (or Wigner) law. In particular,
Wigner has shown that the moments of the limiting distribution $m_{2k}$
are determined by recurrent relations
$$
m_{2k} = v^2 \sum_{j_0}^{k-1} m_{2k-2-2j}\,  m_{2j}, \quad m_0 = 1.
\eqno (1.4)
$$
Obviously, $m_{2k+1} = 0$.

\subsection{Gaussian Unitary (Invariant) Ensemble}

If one considers the ensemble of $N\times N$   hermitian matrices
$$
(H^{(N)})_{ij} = {1\over \sqrt N} h_{ij}, 
\eqno (1.5)
$$
where 
$h_{ij} = \a_{ij} + \I \b_{ij}, i\le j$ are jointly independent 
complex gaussian random variables with zero mean values
and variance $v^2$, the density of the probability distribution of $H^{(N)}=H$ 
can be written in the form
$$
Z_N^{-1} \exp \{ - {N\over 2v^2} \ \T\, H^2\},
\eqno (1.6)
$$
where $Z_N$ is the normalization constant. Note that in this case random
variables $\a$ and $\b$ are jointly independent (for more details, see
section 5 of the present article). 

The probability distribution (1.6) is invariant with respect to the 
unitary transformations. Therefore 
the  random matrix ensemble  (1.5)-(1.6) is referred to
as the Gaussian Unitary Ensemble (GUE) of random matrices.
It plays the central
role in the spectral theory of random matrices (see e.g.
\cite{M}).

It is easy to see that $\{H^{(N)}\}$ verifies the conditions imposed on
$\{W^{(N)}\}$ and therefore the semicircle law (1.2)-(1.3) is valid for GUE.
Moreover, convergence (1.2) holds with probability 1 in this case.
 The first study of the maximal eigenvalue 
$\l_{\max}(H^{(N)}) = \Vert H^{(N)}\Vert $
was carried out by S. Geman \cite{G} for the ensemble of real symmetric matrices
having the probability distribution of the form (1.6). 
This ensemble is known as the 
Gaussian Orthogonal Ensemble (GOE) of random matrices. 
Geman has proved that $\l_{\max}(H^{(N)})$ converges with probability 1 as
$N\to\infty$ to the border $2v$ of the support of $d\s(\l;2v)$.

\section{ Band random matrices}

Given a parameter $b$, 
let us consider hermitian random matrices $H^{(N,b)}$ of the form
$$
H^{(N,b)}_{xy} = h_{xy} \sqrt{\Psi^{(b)}_{xy} }, \quad x,y = 1,\dots,N,
\eqno (2.1)
$$
where $\{h_{xy}, \ x\le y\} $ are the same as in (1.5)
with the law  (1.6) and 
$$
\Psi_{xy}^{(b) } = {1\over b} \psi \left({x-y\over b}\right),
$$
where
$$
\psi(t) = \cases {1, & if $\vert t\vert \le 1/2$, \cr
0, & otherwise\cr }.
\eqno (2.2)
$$

The ensemble of band random matrices (2.1)-(2.2) was considered in relation to
the quantum chaos theory and solid state physics (see \cite{CIM,H,MF}).
The most intriguing question was that the ratio $b^2/ N$ 
separates two major asymptotic regimes that  
characterize   behavior of certain spectral characteristics of random
matrices.

In paper \cite{KP}, the real symmetric analog of (2.1) is considered. It is
proved that in the limit $b\to\infty, b =o(N)$ the Wigner law is valid, i.e.
the normalized eigenvalue counting function 
$
\s_{N,b}(\l) $ of $ H^{(N,b)}$ weakly converges in probability
to the semi-circle distribution (see also papers \cite{CG,KLH,MPK} for this and
related result). 
The same proposition can be easily proved for the
ensemble of hermitian band random matrices. 
If $b= O(N^\gamma)$ with some $\gamma>0$, then the convergence of $\s_{N,b}$ to
the
semicircle distribution  holds with probability 1. 

Much less is known about the limiting behavior of the spectral norm 
$\Vert H^{(N,b)}\Vert =\l_{\max}^{(N,b)}$  of band random matrices.
 Up to our knowledge, 
this question was addressed in paper \cite{BMP} only. 
If one looks at the computations presented in \cite{BMP}, one can easily see
that $\Vert H^{(N,b)}\Vert\to\infty$ as $b = o(\log N), b\to\infty$.
In present paper we show that if $b$ goes to infinity faster than 
$(\log N)^{3}$, then the spectral norm of band random matrices remains 
bounded by $2v$.

Our main result is given by the following proposition.
\vskip 0.5cm

\noindent {\bf Theorem 2.1.}
 
\noindent \textsl{If  $N$ and $b$ go to infinity in the way such that $b/(\log
N)^{3}
\to
\infty$, then the spectral norm
$\Vert H^{(N,b)}\Vert$ remains bounded with probability 1; 
$$
\limsup_{N,b\to\infty } \Vert H^{(N,b)}\Vert  \le  2v.
\eqno (2.4)
$$
If $b=O(N^{\gamma})$ with $\gamma>0$, then (2.4) turns to equality.
}

\vskip 0.5cm

To study the spectral norm of random matrices $H$, 
we employ the general approach used first by
S. Geman by suggestion of U. Grenander \cite{G}.  Later it was employed by many
 authors  in applications to various random matrix ensembles with jointly
independent entries 
\mbox{\cite{BY,FK,K1}} and also for random matrices whose elements are
statistically dependent random variables \mbox{\cite{BK}}. The key
observation is that if one considers the moments 
$$
M_{2k}^{(N)} = \E \left\{{1\over N} \, \T \, [H^{(N)}]^{2k}\right\}
$$ 
in the limit $N\to\infty$, $k= O(\log N)$,
then the leading contribution to $M_{2k}^{(N)}$ will be given by the maximal
eigenvalue of $\l_{\max}(H)$. 

The next  important step  was made in \cite{BS}. 
It was observed that
to study the limiting behavior of the moments $M_{2k}^{(N)}$ of the Wigner
ensemble (1.1), it is not necessary to compute them explicitly. It is
sufficient
to show that they verify a system of recurrent relations that 
converges as $N,k\to\infty$ to the system of equalities (1.4) that determines
the moments of the semicircle law.

In paper \cite{BK} we have developed a method of derivation and study the
recurrent relations for the moments
$M_{2k}$ in the case when the matrix elements are gaussian correlated random
variables. 
In present paper we develop a new version of this method
that we adapt to the band random matrix ensemble.
Also we present more precise analysis of these relations
aiming the best possible estimates of the moments $M_{2k}^{(N,b)}$ of
$H^{(N,b)}$.

We start with the basic example of the GUE (see Section 3).
We pay much attention to this ensemble because the relations and estimates
of the moments $M_{2k}^{(N,b)}$ are very similar to those of GUE.
Thus, the  estimates we get in the GUE case are
true
for the moments of the band random matrices. We did not manage to get
the optimal estimates for $M_{2k}^{(N)}$ obtained by other methods. 
The benefit of the approach developed is that it can be applied for 
random matrix ensembles different from GUE.

In Section 4 
we modify our approach and  study the band random matrices.
These computations lead to the  proof of Theorem 2.1. In Section 5 we prove
auxiliary statements.

\section{Moments of GUE matrices}

\subsection{Recurrent relations }

In this section we derive two main recurrent relations for 
the moments of $H^{(N,b)}$ and their variances.
Let us consider the normalized trace $L_a ={1\over  N} \, \T \, H^a$  
and compute 
the mathematical expectation with respect to the Gaussian measure (1.4)
of the following expression, where $G$ denotes some regular function of $H$:
$$
 \E \{ L_a G\} = {1\over N} \sum_{x,s=1}^N
\E \{ H_{xs} H^{a-1}_{sx} G\}.
$$
Regarding the last average, we can use the 
integration by parts formula
$$
\E \{ H_{xs} H^{a-1}_{sx} G\} = \E \{ \vert H_{xs}\vert^2\} \ 
\E \left\{ { \partial  H^{a-1}_{sx} G\over \partial H_{sx}} \right\}.
\eqno (3.1)
$$
The  use of the partial derivative symbol is 
explained in section 5.

Then we can write that 
$$
\E \{ L_a G\} = { v^2\over N^2} \sum_{x,s=1}^N \sum_{j=0} 
\left[ \E \{H^{a-2-j}_{ss} H^{j}_{xx} R\} + 
\E \{ H^{a-1}_{sx} \partial G/\partial H_{sx} \}
\right] =
$$
$$
v^2 \sum_{j=0}^{a-2} \E \{ L_{a-2-j} L_{j} G\} + 
{v^2\over N^2}\sum_{x,s=1}^N \E \{ H^{a-1}_{sx} \partial G/\partial H_{sx} \}.
$$
If $G=1$ and $a=2k$, then we get the following relation
$$
\E L_{2k}  =
v^2 \sum_{j=0}^{2k-2} \E\left\{ L_{2k-2-j}   L_{j} \right\}.
$$
Denoting $L^o = L - \E L$, we can write that 
$$
\E \left\{ L_{a_1} L_{a_2} \right\} = 
\E L_{a_1} \E L_{a_2} + D_{a_1,a_2}^{(2)}, \quad 
D^{(2)}_{a_1,a_2} = \E \{ L^o_{a_1} L^o_{a_2}\}.
$$
Then
using the fact that 
$M_{2k+1}^{(N)} = \E L_{2k+1} = 0$ (see Section 5), we obtain 
our first  equality
$$
M_{2k}^{(N)} = v^2\sum_{j=0}^{k-1} M_{2k-2-2j}^{(N)} M_{2j}^{(N)} +
v^2\sum_{a_1+a_2 =2k-2} D^{(2)}_{a_1,a_2}
\eqno (3.2)
$$
that form the system of recurrent relations.
Let us note that for two random variables we always have $\E (L_1^o L^o_2) = 
\E (L_1 L^o_2)$. 

Following our general  scheme, we introduce variables
$$
D^{(q)}_{a_1,\dots,a_q} = \E \left\{ L^o_{a_1} L^o_{a_2} \cdots
L^o_{a_q}\right\} =\E \left\{ L_{a_1} [L^o_{a_2} \cdots L^o_{a_q}]^o\right\}
$$
and apply (3.1) to this expression with $G = [L^o_{a_2} \cdots L^o_{a_q}]^o$.
Then we obtain relation
$$
D^{(q)}_{a_1,\dots,a_q} = 
v^2 \sum_{j=0}^{a_1-2} 
\E \left\{ L_{a_1-2-j} L_{j} L^o_{a_2} \cdots L^o_{a_q}\right\} +
$$
$$
{v^2\over N^2}\sum_{i=2}^q 
\E \left\{ L^o_{a_2} \cdots L^o_{a_{i-1}} 
\ a_i \ L_{a_i+a_1-2} \dots L^o_{a_q}\right\}.
$$
The last term arises  because of equality 
$$
\sum_{t=1}^N
{\partial 
(L_a)_{tt} \over \partial W_{sx} } = \sum_{j=0}^{a-1} \sum_{t=1}^N
W^{a-1-j}_{ts} W^j_{xt}= (a-1) \left(W^{a-1}\right)_{xs}.
$$
Now we use two times the identity
$$
\E \{L_1 L_2 Q\} = \E \{L_1 L_2^o Q\} +
\E \{L_1^o L_2 Q\}+
\E \{L_1^o L_2^o Q\}-
\E \{L_1^o L_2^o \} \E \{Q\}
$$
and obtain  our second recurrent relation
$$
D^{(q)}_{a_1,\dots,a_q} = v^2\sum_{j=0}^{a_1-2} M_j^{(N)}
D^{(q)}_{a_1-2-j,a_2,\dots,a_q}+
v^2\sum_{j=0}^{a_1-2} M_{a_1-2-j}^{(N)}
D^{(q)}_{j,a_2,\dots,a_q}+
$$
$$
v^2 \sum_{j=0}^{a_1-2} D^{(q+1)}_{j, a_1-2-j,a_2,\dots,a_q}-
v^2 \sum_{j=0}^{a_1-2} D^{(2)}_{j, a_1-2-j} D^{(q-1)}_{a_2,\dots,a_q}+
$$
$$
{v^2\over N^2} \sum_{i=2}^q a_i M_{a_1 +a_i-2}^{(N)} 
D^{(q-2)}_{a_2,\dots,a_{i-1}, a_{i+1},\dots,a_q} +
{v^2\over N^2}\sum_{i=2}^q a_i D^{(q-1)}_{a_2,\dots,a_{i-1}, a_i+a_1-2,
a_{i+1},\dots,a_q}.
\eqno (3.3)
$$
When deriving (3.3), we assumed  $q>3$. For $q=2$ 
we have equality
$$
D_{a_1,a_2}^{(2)} =
2v^2 \sum_{j=0}^{a_1-2} M_{j}^{(N)} D_{a_1-2-j,a_2} +
v^2 \sum_{j=1}^{a_1-3}D^{(3)}_{j,a_1-2-j,a_2} +{a_2\over N^2}v^2 M_{a_1+a_2-2}
.
\eqno (3.4)
$$
In the case of $q=3$ we  adopt relation (3.3) assuming that $D^{(0)}=1$ and
$D^{(1)}=0$. Also one has to remember that $M_{2l+1}^{(N)} = 0$. 

\subsection{Recurrent estimates}

Let us  prove the following proposition.

\vskip 0.3cm 
\noindent {\bf Lemma 3.1}
\textsl{If $N\ge 4$ and $k\le N^{1/3}$, then }
$$
M_{2k}^{(N)} \le \left(1 + {4k^3\over N^2}\right)^{k}\ m_{2k}
\eqno (3.5)
$$
\textsl{and}
$$
\vert D^{(q)}_{a_1,\dots,a_q}\vert \le 
{a_1 a_2 \cdots a_q\over N^q} \left(1+ {4k^3\over N^2}\right)^{k} \ m_{2k} 
,
\eqno (3.6)
$$
\textsl{where} $ 2k = a_1+\dots+a_q$.

\vskip 0.5cm

We prove this statement following the procedure of recurrent estimates 
proposed in
\cite{BK}.  Let us consider the plane of positive integers, i.e. the set of
points
$B = \{ (l,p): l,p \in {\bf N}\}$. Actually, $B$ is a quarter-plane.

We say that the random variable 
 $D^{(q)}_{a_1,\dots,a_q}$ with given $A_q = (\a_1,\dots, \a_q)$ belongs to
the  point $(2k,q)\in B$, if the sum of $a$'s is equal
to
$2k$.  We call such variables $D$ the elements of $(2k,q)$.
Obviously, $2\le q\le l, l=2k $ and we restrict ourselves with the corresponding
subdomain
$\tilde B$ of $B$.

\begin{figure}[htbp]
\centerline{\includegraphics[width=12cm]{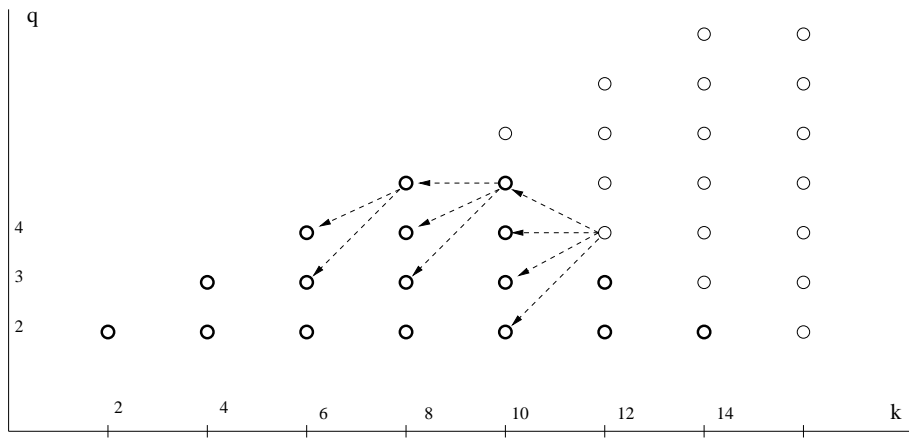}}
\caption{The quarter-plane of recurrent relations for $D$}
\end{figure}

It follows from relations (3.3) that the elements $D$ of $(2k,q)$ are expressed
in terms  of the elements $D$ belonging to one of the four other points: $
(2k-2,q)$,
$(2k-2,q+1)$, $(2k-2,q-1)$ and $(2k-2,q-2)$. 
If any of these points
is situated outside of the domain $\tilde B$, then
the corresponding term in (3.3) is equal to zero and gives no contribution. 
On figure 1 we represent three possible situations. The fact that
the element of $(2k,q)$ is expressed in terms of other elements is reflected
by a flesh. Let us note that the right-hand side of (3.3) uses also 
the moments $M_{2l}$ with $2l\le 2k-2$. So, the flesh lines 
directed from points $(2k,\cdot)$ to $(2k-2,\cdot)$
also indicate the dependence of $D$ on $M$.

Suppose we want to prove the estimate (3.6) for the elements $D$ 
of the point
$(2k',q')$ with given $k'$ and $q'$. On Figure 1 this is the point (10,5).
We assume that these estimates are true for the elements $D$ of all points
$(2k'',q'')$ that belong to the triangle $2k''+q'' < 2k'+q'$ 
and of those situated on the line 
$2k''+q'' = 2k'+q'$ with $2k''\le 2k'-2$. On Figure 1 we marked these points by fat circles.
Also we assume that (3.5) holds for all $M_{2l}$ with $2l\le 2k'-2 +q'$. 
Then we use relations (3.3) to deduce the estimate (3.6) 
for the point $(2k',q')$.

So, starting with the point $(2,2)$, we obtain the estimates
(3.6)  for the elements of $(4,3)$ and $(4,2)$. Then  
we go step by step  over the points of $\tilde B$  until  we reach the
point $(2k',q') = (10,5)$.
The next points are (12,4), (14,3) and so on.
On this way we proceed along the line
$2k''+q''= 2k'+q'$ until $q''=2$ (the point (16,2) on Figure 1). Using (3.2),
we prove the estimate (3.2)  for $M_{2k'+q'}$. Then we complete the next line
starting from the point (10,6).

Computationally this recurrent procedure means that we assume that 
all terms of the right-hand side of (3.3) verify inequalities (3.5) and
(3.6) and then estimate their sum to show that it is less than the
expression corresponding to the term of  the left-hand side of (3.3). 
The same is true for (3.2) with the initial point of the recurrence
$M_{0}^{(N)} = 1$. We note also that $M_{2}^{(N)} = v^2$
that obviously verifies (3.5) with $k=1$ because $m_2 = v^2$.
Let us have a look at $D_{1,1}^{(2)}$ that serves as the initial point for
(3.3). It is easy to see that
$$
D_{1,1}^{(2)} = \E \left\{ {1\over N} \T\,  H \  Ê{1\over N} \T\,  H\right\} =
{v^2\over N^2}
\eqno (3.7)
$$
that certainly satisfies (3.6).

Now let us  consider (3.2). We want to show that if (3.5) and (3.6)
are valid for the terms of the right-hand side of (3.2), then 
(3.5) is valid for $M_{2k}^{(N)}$.
The first term of the right-hand side of (3.2) is bounded by
$$
v^2\sum_{j=0}^{k-1} \left(1+{4(k-1-j)^3\over N^2} \right)^{k-1-j}
\left(1+{4j^3\over N^2}\right)^{j}\ m_{2k-2-2j} m_{2j} + 
\le 
$$
$$
\left(1 + {4(k-1)^3\over N^2}\right)^{k-1} 
 v^2\sum_{j=0}^{k-1}m_{2k-2-2j}
m_{2j} \le
 \left(1 + {4k^3\over N^2}\right)^{k-1} m_{2k}.
$$

The second term of the right-hand side of (3.2)
is bounded by
$$
v^2 m_{2k-2} \left(1 + {4(k-1)^3\over N^2}\right)^{k-1} 
\sum_{a_1+a_2 = 2k-2} {a_1 a_2\over N^2} \le
\left(1 + {4k^3\over N^2}\right)^{k-1} {(2k-2)^3\over 2N^2} m_{2k}
$$
because
$$
\sum_{j=1}^{2k-3} (2k-2-j)j = \sum_{i=1}^{k-1} (2i-1)^2 < (2k-2)^3/2.
$$
Gathering these two estimates, one obtains that
$$
M_{2k}^{(N)} \le  \left(1 + {4k^3\over N^2}\right)^{k-1} \left[ 1+
{4(k-1)^3\over N^2}\right] m_{2k}
$$
and (3.5) obviously follows.

Now let us turn to (3.3). Let us denote
$$
{a_1 a_2 \cdots a_q\over N^q} \left(1+ {4k^3\over N^2}\right)^{k} 
\equiv  \Pi(A_q;N). 
$$
 We assume that (3.5) and (3.6)
are valid for the all terms of the right-hand side of (3.4) and 
show that their sum is bounded by the expression standing at the
right-hand side of (3.6).
Let us consider two first terms of the right-hand side of (3.4). 
Repeating calculations that lead to  (3.7), we can write that the sum of the
first  two terms
is estimated by
$$
{v^2a_2\cdots a_q\over N^q} \left( 1+ {4k^3\over N^2}\right)^{k-1}
\left[
\sum_{j=1}^{a_1-3} (a_1-2-j)  \ m_j 
\ m_{2k- 2 -j} +\sum_{j=1}^{a_1-3} j  \ m_{a_1-2-j}
\ m_{2k -a_1 +j}
\right]
$$
The terms in square brackets represent
 the first and the third parts of the sum 
$$
\left[(a_1-3) \cdot m_1 m_{2k-3} + \dots + 2 m_{a_1-4} m_{2k-a_1+2} +1\cdot
m_{a_1-3} m_{2k-a_1+1}\right] +
$$
$$
\left\{m_{a_1-2} m_{2k-a_1} + m_{a_1-1} m_{2k-a_1-1} + \dots + 
m_{2k-a_1} m_{a_1-2}\right\}+
$$
$$
\left[1\cdot m_{2k-a_1+1} m_{a_1-3} + 2  m_{2k-a_1+2}m_{a_1-4}+\dots  +(a_1-3)
m_{2k-3} m_{1}\right]\le 
$$
$$
(a_1-3) \sum_{j'=1}^{2k-3} m_{j'}m_{2k-2-j'}.
$$
Certainly,  all the sums run over even numbers,
but we do not care about this because $m_{2j+1} = 0$ and all relations are
still true when regarding the sums over even and odd numbers. 
Thus we obtain that the sum of two first terms of the right-hand side of (3.4)
is bounded by
$$
\Pi(A_q;N)
\left(1 - {3\over a_1}\right) m_{2k}.
\eqno (3.8)
$$
Now let us consider the third term of the right-hand side of (3.4). Assuming
that (3.5) and (3.6) hold for
corresponding
$M$ and $D$, we can write that
$$
v^2\sum_{j=1}^{a_1-3}\vert  D^{(q+1)}(j, a_1-2-j,a_2,\dots,a_q) \vert \le
$$
$$
\Pi(A_q;N) {v^2\over N} m_{2k-2}  \sum_{j=1}^{a_1-3}
{ j(a_1-2-j)\over a_1}\le 
{\Pi(A_q;N)}  {
a_1^2\over 2 N} m_{2k}.
\eqno (3.9)
$$
The fourth term of the right-hand side of (3.4) is estimated 
 by the same expression.

The fifths term is estimated by 
$$
{v^2a_2\cdots a_q\over  N^{q+1} } \sum_{i=2}^q 
\left(1+{({a_1+a_i}-2)^3\over 2N^2}\right)^{{a_1+a_i-2\over 2}}
\left(1+{(2k-a_1-a_i)^3\over 2N^2}\right)^{{2k-a_1-a_i\over 2}} m_{a_1+a_i-2}
m_{2k-a_1-a_i}\le
$$
$$
\Pi(A_q;N){1\over a_1} 
v^2 \sum_{i=2}^q  m_{a_1+a_i-2}
m_{2k-a_1-a_i}\le 
\Pi(A_q;N)  {1\over a_1} m_{2k}.
\eqno (3.10)
$$
Finally, the last term of (3.4) is less or equal to
$$
{v^2\over N^{q+1}} \sum_{i=2}^q {(a_1+a_i-2)\ a_2\cdots a_q}\left(1+
{4(k-1)^3\over N^2}\right)^{k-1} m_{2k-2} \le
$$
$$
\Pi(A_q;N) 
v^2 m_{2k-2} \sum_{i=2}^q {a_1+a_i -2\over a_1 N}\le 
 \Pi(A_q;N) {2k-2\over N} 
m_{2k}.
\eqno (3.11)
$$
The sum of all expressions (3.6)-(3.9) gives us inequality
$$
\vert D^{(q)}_{a_1,\dots, a_q} \vert \le \Pi(A_q;N) \left[ 
\left(1-{2\over a_1}\right) + { a_1^2\over N} + {2k-2\over N}
\right]m_{2k}.
$$
Taking into account that $a_1 \le 2k-2$, it is easy to show that the
sum of the terms in square brackets is strictly less than $1$ provided $k\le
N^{1/3}$. This implies inequality (3.6).

Lemma 3.1 is proved.

\subsection{Asymptotic behavior of the moments of GUE matrices}

\subsubsection{Convergence of the maximal eigenvalue}
It is very well known  the numbers $m_{2k} =
m_{2k}(2v)$ determined by (3.12) given by Catalan numbers
$$
m_{2k} = v^{2k} {1\over k+1} {2k \choose k}.
\eqno (3.12)
$$
Elementary computations imply that $m_{2k} (2v) \le (2v)^{2k}$. Then it
follows from Lemma 3.1 that given $\vep >0$, we have estimate
$$
M_{2k}^{(N)} \le \left(2v \sqrt{1+\vep}\right)^{2k} 
\eqno (3.13)
$$
for all $k\le N^{1/3}$ provided $4k^3/N^2 \le \vep $.

Taking into account inequality $(\l_{\max})^{2k} \le \sum_{i=1}^N \l_i^{2k} =
{\hbox{Tr}} H^{2k}$, we can write that 
$$
{\hbox{Prob}}\{\l_{\max} \ge 2v(1+\vep)\} \le
{N M_{2k}^{(N)}\over \left[ 2v (1+\vep)\right]^{2k}}\le 
 {N\over  (1+\vep)^k }
\quad {\hbox{ for all }} \ k\le N^{1/3}.
\eqno (3.14)
$$
Regarding $k\gg \log N$ and using the Borel-Cantelli lemma, it is easy to 
 deduce from (3.14) that 
$\limsup_{N\to\infty} \l_{\max}^{{{(GUE)}}} \le 2v$
with probability 1.
This estimate together with the convergence
(1.2) with probability 1 implies that
$\l_{\max}\to 2v$ 
 with probability 1 as $N\to\infty$.

\subsubsection{High moments and the scale at the spectral edge}

Let us compare our results with those already known for the moments of GUE
with $v^2=1/4$.
It is known that the moments $M_{2k}^{(N)}$ are given by the following
 recurrent relation \cite{HZ}:
$$
M_{2k}^{(N)} = {2k-1\over 2k+2} \ M_{2k-2}^{(N)} +
{2k-1\over 2k+2} \cdot {2k-3\over 2k}\cdot  {k(k-1)\over 4N^2}\ 
M_{2k-4}^{(N)}.
\eqno (3.15)
$$
Explicit expression for $m'_{2k} = m_{2k}(1)$
(3.12) implies that  
$$
m'_{2k} = {2k-1\over 2k+2} m'_{2k-2} = {2k-1\over 2k+2} \cdot 
{2k-3 \over 2k} m'_{2k-4}. 
$$
Then it is  easy to 
deduce from (3.15) that
$M_{2k}^{(N)}$ admit the following estimates (see, for example \cite{L}):
$$
M_{2k}^{(N)} \le \left(1 + {k^3\over 4N^2}\right) m'_{2k} \quad {\hbox{for
all\ \ }} k,N.
\eqno (3.16)
$$
Also one can write that
$$
M_{2k}^{(N)} \le \left(1 + {k^2\over 4N^2}\right)^k m'_{2k} \quad {\hbox{for
all\ \ }} k,N.
\eqno (3.17)
$$
Inequality (3.16) means that the moments $M_{2k}^{(N)}$ admit
the power-like estimates by $(2v)^{2k}(1+o(1))$ in the limit
$1\ll k \ll N^{2/3}$ and that this behavior can change on the regime
$k = t N^{2/3}$.

Inequality (3.16) can be obtained by using the orthogonal polynomial
approach (see the early paper by Bronk \cite{B}, where  the
scaling at the edge  of the semicircle distribution 
has been determined for ther first time); explicit
asymptotic expressions  were  found in the seminal paper by Tracy and 
Widom
\cite{TW}. It was  shown that the fraction $k^3/N^2$ is really the optimal
one in the sense that one cannot decrease the exponent of $k$ and increase that
of $N$  in (3.16).

Basing on the relations (3.2) and (3.3), we did not manage
to obtain estimates as precise as (3.16).
Our result (3.5) implies that 
$$
M_{2k}^{(N)} \le \left(1 + C{k^4\over N^2}\right) m_{2k} \quad {\hbox{for \ \
}}
k\le N^{1/3}.
\eqno (3.18)
$$
Inequalities of this type are sufficiently powerful to 
estimate  the maximal eigenvalue of
$H$ but they do not reflect the real scale of the eigenvalue distribution
at the edge of the limiting spectrum.

From another hand, we did not used the orthogonal polynomial approach
to obtain (3.2) and (3.4). Therefore the positive counterpart is that
our approach is applicable for more general ensembles of random matrices
than the GUE. In the next section we show how our approach 
works in the case of band random
matrices. 

\section{Moments of band random matrices}

\subsection{Main technical proposition and proof of Theorem 2.1}

Our main goal is to study 
the moments
$$
M_{k}^{(N)} = \E\left\{{1\over N} {\hbox{Tr }} [H^{(N,b)}]^{k}\right\},
\eqno (4.1)
$$
where $H^{(N,b)}$ are given by matrices (2.1). 
The first observation is that $M_{2k+1}^{(N,b)} = 0$ (see Lemma 5.2 of
section 5). Thus we can consider the even moments
$M_{2k}^{(N,b)}$ only.
The next observation is that in the case of band random matrices
we cannot derive recurrent relations for $M_{2k}$ themselves as it was for the
case of GUE. 
Instead we find a family of random variables 
that make a close system. 
Let us start with the random variables
$$
L_k(x) = L_x^{(N,b)} = [H^{(N,b)}]^{k}_{xx}.
$$
Using integration by parts formula (see section 5),
it is not hard to show that the mathematical expectation of $L$ 
verifies the following identity
$$
\E L_{2k}(x) = v^2 \sum_{j=0}^{2k-1}\E\{ L_{2k-2-j}(x) L_{j}[x]\},
\eqno (4.2)
$$
where we denoted 
$$
L_j[x] = {1\over b}\sum_{s=1}^N L_j(s) \ \psi\left({s-x\over b}\right).
$$
One can say that $L_j[x]$  
represents a partial trace of $H^j$ 
normalized by $b$. So, one can expect that the variance of $L$ goes to zero
when
$b\to\infty$ and that the mathematical expectation in the right-hand side of
(4.2) factorizes. In what follows, we prove this factorization and estimate
the variance of $L[x]$.

Regarding mathematical expectation
$ M_{2k}^{(N,b)}(x) = \E L_x(k)$, we obtain equality 
$$
M_{2k}^{(N,b)}(x) = v^2 \sum_{j=0}^{k-1} M_{2k-2-2j}^{(N,b)}(x) \
M_{2j}^{(N,b)} [x] + v^2 D^{(2)}_{2k-2}(x),
\eqno (4.3)
$$
where 
$$
D^{(2)}_{2k-2}(x) = \sum_{a_1 + a_2 = 2k-2} \E \{L^o_{a_1}(x)
L^o_{a_2}[x]\}.
$$
We recall that $L^o = L - \E L$.

In the next subsection we will prove Lemma 4.1 that implies the following
estimates that are true for all $ 2k\le b^{1/3}$:
$$
\sup_{x = 1,\dots,N} M_{2k}^{(N,b)}(x) \le \left(1+{4k^3\over
b^2}\right)^{k} m_{2k},
\eqno (4.4)
$$
where the family $\{m_{2k}\}_{k\in {\bf N}}$ determines the semicircle law
(1.2).
\vskip 0.5cm

It follows from (4.4) that
$$
M_{2k}^{(N,b)} = {1\over N} \sum_{x=1}^N M_{2k}^{(N)}(x) 
\le \left(1+{4k^3\over
b^2}\right)^{k} m_{2k}
\eqno (4.5)
$$
Using the estimate (3.14),
we see that 
$$
{\hbox{Prob}}\{ \l_{\max}^{(N,b)} \ge 2v(1+\vep) \} \le {N\over(1+\vep)^k}
$$
for all $k\le b^{1/3}$ provided $4k^3/b^2 \le \vep$. 
Choosing $k$ such that $k/\log N \to \infty$,
we obtain the bound
$$
\limsup_{N\to\infty} \max_{j=1,\dots N} \vert \l^{(N,b)}\vert \le 2v 
\eqno (4.6)
$$
with probability 1 provided $b\ge (\log N)^3$.
Theorem 2.1 follows form (4.6).

\subsection{Recurrent relations for the generalized moments}

Let us introduce random variables that serve as the elementary blocks 
for the closed system of recurrent relations. These are the products
$$
L_{R(a)}(x) = \left( H^{\a_1} \Psi^{(y_1)} H^{\a_2} \cdots H^{\a_{r-1}}
\Psi^{(y_{r-1})} H^{\a_r}\right)_{xx},
$$
where $R$ is given by two $r$-dimensional 
vectors $A$ and $Y$; $A_r =(\a_1,\dots,\a_r)$, 
$\a_1+\dots+\a_r = a$,    $Y_r = (y_1,\dots,y_r)$, $y_i\in \{1,\dots,N\}$,
and $\Psi^{(y)}$ represents a diagonal $N$-dimensional matrix
$[\Psi^{(y)}]_{st} = \delta_{st}\psi((s-y)/b)$. 

In section 5 (see Lemma 5.2)  we prove that 
$$
\E L_{R(a)} = 0 \quad {{\hbox{ if \ \ }}} a=2k+1.
\eqno (4.7)
$$
In present section we prove the following proposition.
\vskip 0.5cm 
{\bf Lemma 4.1} \textsl{If $2k\le b^{1/3}$, then 
$$
0\le \E L_{R(a)}\le  \left(1+ {\displaystyle 4k^3\over
\displaystyle b^2}\right)^{k} m_{2k} \quad {{\hbox{ if \ \ }}} 
a=2k,
\eqno (4.8)
$$
and 
$$
\sup_{x,y}\E\left\{ L^o_{R_1(a_1)}(y) \ L^o_{R_2(a_2)} [x] \cdots
L^o_{R_q(a_q)}[x]\right\}\le 
\cases{  \Pi (A_q;b) \ m_{2k},
& if $\vert A_q\vert  = 2k$,\cr
0, & if $\vert A_q\vert  = 2k+1$,\cr}
\eqno (4.9)
$$
where
$$
\Pi(A_q;b) = 
{a_1a_2\cdots a_q\over b^q} 
\left(1+{4k^3\over b^2}\right)^{k}
$$
and}
$$
\vert A_q\vert = a_1+\dots +a_q.
$$

\vskip 1cm

To prove estimates (4.8) and (4.9), we derive a system of recurrent
relations
that resemble very much  relations (3.2), (3.3),  and  
(4.3). Then we use the recurrent procedure described in subsection 3.2.
So, we do not describe the details of the derivation but give the general 
description. 

Regarding $L_{R(a)}$, one can always assume that $\a_1>1$. Therefore one can write
that 
$$
\E\left\{ \sum_{s=1}^N H_{xs}\left( H^{\a_1-1}\Psi_1 \cdots 
\Psi_{r-1} H^{\a_r}\right)_{sx}\right\} = 
$$
$$
v^2\sum_{l=1}^r \sum_{j=0}^{\a'_j-1} 
\E\left\{\sum_{s=1}^N 
(H^{\a_1-1}\Psi_1\dots \Psi_{l-1} H^j)_{ss} {1\over b} \psi({s-x\over b})
(H^{\a_l-1-j}\Psi_{l} \cdots  \Psi_{r-1} H^{\a_r})_{xx}\right\},
$$
where we denoted $\Psi_i= \Psi^{(y_i)}$ and
$$
\a_l' = \cases{\a_1 - 2, & if $l=1$\cr
\a_l-1, & if $l\neq 1$.\cr}
$$ 
Now, factorizing the mathematical expectation,
one gets
$$
\E L_{R(a)}(x) = 
v^2\sum_{l=1}^r \sum_{j=0}^{\a'_j-1} 
\E\left\{
(H^{\a_1-1}\Psi_1\dots \Psi_{l-1} H^j)[x]\right\}
\E \left\{
(H^{\a_l-1-j}\Psi_{l} \cdots  \Psi_{r-1} H^{\a_r})_{xx}\right\}+
$$
$$
v^2\sum_{l=1}^r \sum_{j=0}^{\a'_j-1} 
\E\left\{
(H^{\a_1-1}\Psi_1\dots \Psi_{l-1} H^j)^o[x]
(H^{\a_l-1-j}\Psi_{l} \cdots  \Psi_{r-1} H^{\a_r})^o_{xx}\right\}.
\eqno (4.9)
$$
If one accepts the symbolic denotation $\sum_{l=1}^r \sum_{j=0}^{\a'_j-1} \equiv 
\sum_{J=0}^{\vert A\vert-2 }$, then
we can rewrite (4.9) in the form close to (4.2)
$$
\E L_{R(a)}(x) = \sum_{J=0}^{a-2} E\{ L_{J}[x]\} 
\E\{ L_{a-2-J}(x)\} + \sum_{J=0}^{a-2} \E \{L^o_J[x] L^o_{a-2-J}(x)\}.
$$

Now let us derive recurrent relations for $D$; to simplify the formulas,
we accept denotation
$D^{(q)}(y,x)= \E \{ L_1^o(y) L_2^o[x]\cdots L^o_q[x]\}$ with obvious agreement
that the vectors $A_i,Y_i$ that correspond to $L_i$ are  not necessarily the same
for different $i$. 
Denoting $T_{2,\dots,q}= L_2^o\cdots L_q^o$ and mimicking 
computations of section 3, and those presented above, we obtain that
$$
D^{(q)}(y,x) = v^2 \sum_{J=1}^{a_1-3} 
\E\{ L_J[x] L_{a_1-2-J} T^o_{2,\dots,q}\} +
$$
$$
{v^2\over b^2 } 
\sum_{s,t=1}^N  \Psi_s^{(y)} \Psi_t^{(x)} \sum_{p=2}^q
\sum_{l=1}^p \sum_{j=0}^{\a_l^{(p)}-1} 
\E\{ L_2^o \cdots L_{p-1}^o 
\left( H^{\a_1^{(p)}}\Psi \cdots \Psi H^{\a_l^{(p)}-1-j}\right)_{ts}
\times
$$
$$
\left(H^j\Psi\cdots \Psi H^{\a_{r_p}^{(p)}}\right)_{yt}
L_{p+1}^o\cdots  L_q^o \left(H^{\a_1^{(1)}-1}\Psi 
\cdots \Psi H^{\a^{(1)}_r}\right)_{sy}.
\eqno (4.10)
$$
After factorization of the mathematical expectation, the first term
of the right-hand side of (4.9) produces four terms
$$
\E\{ L_J^o[x]L_2^o[x]\cdots L_q^o[x]\} \E\{ L_{a_1-2-J}(y)\}+
\E\{ L_{a_1-2-J}^o(y)L_2^o[x]\cdots L_q^o[x]\} \E\{ L_{J}[x]\}+
$$
$$
\E\{ L_{a_1-2-J}^o(y)L_J^o[x]L_2^o[x]\cdots L_q^o[x]\} - 
\E\{ L_2^o[x]\cdots L_q^o[x]\} \E\{ L_{a_1-2-J}^o(y)L_J^o[x]\}.
\eqno (4.11)
$$
The estimates of corresponding sums $v^2\sum_{J=1}^{a_1-3}$ are the same as
those obtained in section 3 with the only difference that we take $\sup_{x,y}$
of the left-hand sides and replace in (3.8) and (3.9) 
the factor $\Pi (A_q;N)$ by $\Pi(A_q;b)$.

Let us consider the last term of (4.9).
As one can see, it produces the terms of the forms ${1\over b^2}M D^{(q-2)}$
and ${1\over b^2} D^{(q-1)}$.
Let us consider them  in more details. 
For the first term we can write that
$$
{v^2\over b^2} \sum_{p=2}^q \sum_{l=2}^p\sum_{j=0}^{\a_l^{(p)}-1}
\E\{ L_2^o\cdots L_{p-1}^o L_{p+1}^o\cdots L_q^o\}\times 
$$
$$
\E\{ [H^j\Psi\cdots 
\Psi H^{\a_{r_p}^{(p)}}\Psi^{(x)}H^{\a_1^{(p)}}\Psi\cdots \Psi 
H^{\a_l^{(p)}-1-j}\Psi^{(y)}H^{\a_1^{(1)}-1}\Psi
\cdots\Psi H^{\a_{r_1}^{(1)}}]_{yy}
\}\le 
$$
$$
{v^2\over b^{q+1}}  
\sum_{p=2}^q \sum_{l=2}^p\sum_{j=0}^{\a_l^{(p)}-1}
{a_2\cdots a_q\over a_p} \left(1+{(\vert A_q\vert - a_1-a_p)^3\over
b^2}\right)^ {\vert A_q\vert - a_1-a_q}
\left(1+{(a_1+a_q-2)^3\over b^2}\right)^{a_1+a_q-2}\times
$$
$$
m_{\vert A_q\vert -a_1-a_p} m_{a_1+a_p-2}\le 
\Pi(A_q;b) {1\over a_1} m_{2k}.
\eqno (4.12)
$$
Here we have used the fact that
$\sum_{l=2}^p\sum_{j=0}^{\a_l^{(p)}-1} 1 = a_p-1$.
The remaining estimate is
$$
{v^2\over b^2} \sum_{p=2}^q \sum_{l=2}^p\sum_{j=0}^{\a_l^{(p)}-1}
\E\{ L_2^o\cdots L_{p-1}^o 
[H^j\cdots 
 H^{\a_{r_p}^{(p)}}\Psi^{(x)}H^{\a_1^{(p)}}\cdots 
H^{\a_l^{(p)}-1-j}\Psi^{(y)}H^{\a_1^{(1)}-1}
\cdots H^{\a_{r_1}^{(1)}}]^o_{yy}
L_{p+1}^o\cdots L_q^o\}\le 
$$
$$
\Pi(A_q;b) \sum_{p=2}^q {a_1+a_p-2\over a_1} \left(1+{(2k-2)^3\over
b^2}\right)^{-2} m_{2k}.
\eqno (4.13)
$$

Now it is easy to gather the estimates of (4.11) with expressions (4.12) and
(4.13) and obtain inequality
$$
\sup_{x,y} \vert D^{(q)} (x,y) \vert \le \Pi(A_q;b) \left[ 
\left(1-{2\over a_1}\right) + { a_1^2\over b} + {2k-2\over b}
\right]m_{2k}.
$$
To complete  the proof of Lemma 4.1, it remains to check that inequalities
(4.8) and (4.9) are true for $L_{R(a)}$ and $D^{(2)}$ (see (3.7)). 

\section{Auxiliary propositions}

\subsection{Integration by parts formula for GUE}

Let us consider random matrices
$$
H_{ij} = \a_{ij} + \i \b_{ij}, \quad 
$$
where $\a$ and $\b$ are real independent gaussian random variables with zero
mean value and variances 
$$
\E \a_{ij}^2 = v^2/2, \quad \E \b_{ij}^2 = v^2/2.
\eqno (5.1)
$$
We consider also the symmetric continuation $A$:
$$
A_{ij} = \cases{\a_{ij},& if $ i\le j$, \cr
\a_{ji}, & if $i> j$ \cr}
$$
and anti-symmetric continuation of $B$.
Then $H = A+ \i B$ is hermitian matrix with the probability distribution
(1.6).
Integration by parts formula implies that
$$
\E H_{xy} G(H) = \E A_{xy}^2 \ \E {\partial G(H) \over \partial A_{xy}}
+ \i \E B_{xy}^2 \ \E {\partial G(H)\over B_{xy}}.
\eqno (5.2)
$$
Let us consider $G(H) = (H^l)_{st}$. Then we can write that
$$
{\partial G\over \partial A_{xy}} =
\lim_{\D \to 0} {1\over \Delta} \sum_{j=1}^{l} \sum_{u,v=1}^N
(H^{j-1})_{su} \left[ \d_{ux}\d_{vy} \D + \d_{uy}\d_{vx} \D\right]
(H^{l-j})_{vt} =
$$
$$
\sum_{j=1}^{l} \left\{
(H^{j-1})_{sx}  (H^{l-j})_{yt} + (H^j)_{sy}  (H^{l-j})_{xt}\right\}.
$$
Similarly we get
$$
{\partial G\over \partial B_{xy}} = 
\i \sum_{j=1}^{l} \left\{
(H^{j-1})_{sx}  (H^{l-j})_{yt} - (H^{j-1})_{sy}  \ (H^{l-j})_{xt}\right\}.
$$
Substituting these relations into (5.2) and remembering (5.1),
we obtain that
$$
\E \left\{ H_{xy} (H^l)_{st}\right\} = 
\E \vert H_{xy}\vert^2 \  \E \left\{ \sum_{j=1}^{l} (H^{j-1})_{sy}
\ (H^{l-j})_{xt}
\right\}.
\eqno(5.3)
$$
Regarding this formula, one can say that formally
$$
\E \left\{ H_{xy} (H^l)_{st}\right\} = 
\E \vert H_{xy}\vert^2 \  \E \left\{ {\partial (H^l)_{st} \over \partial
H_{yx}}
\right\}.
$$
That is the way that we use the partial derivatives in sections 3 and 4.

\subsection{Even and odd moments}

To prove equality (4.7), we consider the mathematical expectation of the
product
$$
\E \left\{ 
H_{xs_1}\cdots H_{s_{\a_1}t_0} \Psi^{(1)}_{t_0} H_{t_0t_1} \cdots 
H_{t_{\a_2}u_0}
\Psi^{(2)}_{u_0} \cdots H_{z_0 z_1}\cdots H_{z_{\a_r}x}
\right\}
\eqno (5.4)
$$
for fixed values of $s_i,t_j, \dots$ and odd sum of $\a_i$'s. 
It is clear that the number of factors $H$ is odd and since we have gaussian
random variables, the mathematical expectation is equal to zero.
The same reasoning shows that the left inequality of (4.8) is true.


\begin{thebibliography}{99}

\bibitem{BDK}
J. Baik, P. Deift, K. Johansson,
{ On the distribution of the length of the longest increasing subsequence of
random permutations},  {\it J. Amer. Math. Soc.}
{\bf  12} (1999) 1119--1178



\bibitem{BY} Z.D. Bai, Y.Q. Yin. Necessary and sufficient conditions 
for almost sure convergence of the largest eigenvalue of
a Wigner matrix,   
{\it    Ann. Probab.}   {\bf 16} (1988) 1729-1741




\bibitem{BMP} L. Bogachev, S.A.  Molchanov, 
L.A. Pastur. On the level density of band random matrices,
  {\it  Math. Notes}   {\bf 50} (1991) 1232-1242





\bibitem{BK} A. Boutet de Monvel, A. Khorunzhy.
On the norm and eigenvalue distribution
of large random matrices,  {\it Ann. Probab. } 
  {\bf 27} (1999) 913-944


\bibitem{BS} A. Boutet de Monvel, M. Shcherbina.
On the norm of random matrices,  
{\it Math. Notes} {\bf  57} (1995), 475--484 
  

\bibitem{B} B.V. Bronk. Accuracy of the semicircle approximation
for the density of eigenvalues of random matrices,
{\it J. Math. Phys.} {\bf 5} (1964) 215-220



\bibitem{CG} G. Casati, V. Girko. Wigner's semicircle law for band random
matrices,  
{\it Rand. Oper. Stoch. Equations   }    {\bf 1} (1993) 15-21

\bibitem{CIM} G. Casati,  L. Molinari, F. Izrailev.
Scaling properties of band random matrices,  
{\it Phys. Rev. Lett.   }   {\bf 64} (1990) 1851


\bibitem{FK} Z. F\"uredi, J. Koml\'os. The eigenvalues of random symmetric
matrices,  {\it  Combinatorica  }   {\bf 1} (1981) 233-241


\bibitem{MF} Y.V. Fyodorov, A.D. Mirlin.
Scaling properties of localization in random band matrices:
a $\s$-model approach,   {\it Phys. Rev. Lett}  
{\bf 67} (1991) 2405


\bibitem{G}  S. Geman. A limit theorem for the norm of random matrices,
 {\it    Ann. Probab.}   {\bf 8} (1980) 252-261 

\bibitem{H} F. Haake, {\it Quantum signatures of chaos.} 
 Springer Series in Synergetics, 54. Springer-Verlag, Berlin, 1991.



\bibitem{HZ} J. Harer,  D. Zagier. The Euler characteristic of the moduli
space of curves. {\it Invent. Math.}  {\bf 85} (1986) 457--485

\bibitem{J} ÊK. Johansson. The longest increasing subsequence in a random
permutation and a unitary random matrix model.
 {\it Math. Res. Lett.} {\bf 5}  (1998)  63--82 


\bibitem{K1} A. Khorunzhy. Sparse random matrices: spectral edge and
statistics
of rooted trees, {\it Adv. Appl. Probab.}   {\bf 33} (2001) 124-140

\bibitem{KK} A. Khorunzhy, W. Kirsch. On asymptotic expansions and scales of
spectral universality in band random matrix ensembles,  
{\it Commun. Math. Phys.}   {\bf 231} (2002) 223-255


\bibitem{KP} A. Khorunzhy, L. Pastur. Limits of infinite
interaction radius, dimensionality and number of components for random
operators with off-diagonal randomness,  {\it Commun. Math. Phys.   }  
 {\bf 153} (1993) 605-646

\bibitem{KLH} M. Ku\'s, M. Lewenstein,  F. Haake.  Density of eigenvalues of
random band matrices. {\it Phys. Rev. A}  {\bf 44} 
(1991) 2800--2808



\bibitem{L} M.  Ledoux.  A remark on hypercontractivity and tail inequalities
for the largest eigenvalues of random matrices.  {\it S\'eminaire de
Probabilit\'es, XXXVIIs}, Lecture Notes in Mathematics,   {\bf 1832} (2003) 
360-369 



\bibitem{M}  M. L. Mehta.   {\it Random Matrices}, Academic Press  (1991)  {\bf
}





\bibitem{MPK} S.A. Molchanov, L.A. Pastur, A.M. Khorunzhy.
Eigenvalue distribution for band random matrices in the limit
of their infinite rank,  {\it  Theoret. and Math. Phys.} {\bf 90}
(1992) 108--118   

\bibitem{TWa} T. Nagao, M. Wadati. Correlation functions of random
matrix ensembles related to classical orthogonal polynomials, 
{\it J. Phys. Soc.
Japan} {\bf  60} (1991)  3298--3322


\bibitem{TW} C.A. Tracy, H. Widom.
Level-spacing distributions and the Airy kernel, 
{\it Comm. Math. Phys.} {\bf 159} (1994) 151--174 


\bibitem{W} E. Wigner. Characteristic vectors of bordered matrices
of infinite dimensions.   {\it  Ann. Math.   } 
{\bf 62} (1955)



















\end{thebibliography}
\end{document}